\documentclass[oneside,english,reqno]{amsart}
\usepackage[T1]{fontenc}
\usepackage[latin9]{inputenc}
\usepackage{geometry}
\usepackage{url}
\usepackage{amsthm}
\usepackage{amssymb}
\usepackage{esint}

\makeatletter
\numberwithin{equation}{section}
\numberwithin{figure}{section}
\theoremstyle{plain}
\newtheorem{thm}{\protect\theoremname}
  \theoremstyle{definition}
  \newtheorem{defn}[thm]{\protect\definitionname}
  \theoremstyle{plain}
  \newtheorem{prop}[thm]{\protect\propositionname}
  \theoremstyle{plain}
  \newtheorem{lem}[thm]{\protect\lemmaname}

\makeatother

\usepackage{babel}
  \providecommand{\definitionname}{Definition}
  \providecommand{\lemmaname}{Lemma}
  \providecommand{\propositionname}{Proposition}
\providecommand{\theoremname}{Theorem}

\begin{document}

\title[Ground-State Wave Function of the One-Dimensional Polaron]{On the Ground-State Wave Function of the One-Dimensional Polaron
in the Strong-Coupling Limit}

\thanks{\copyright 2015 by the author. This paper may be reproduced, in
its entirety, for non-commercial purposes. }

\address{Georgia Institute of Technology, School of Mathematics, Atlanta,
Georgia 30332-0160}

\email{ghanta@gatech.edu}

\author{Rohan Ghanta}
\begin{abstract}
We consider the one-dimensional Fröhlich polaron localized in a symmetric
decreasing electric potential. It is known that the non-linear Pekar
functional corresponding to our model admits a unique minimizer. In
the strong-coupling limit, we show that any approximate ground-state
wave function of our model- after integrating out its phonon modes-
converges in the weak sense to this unique minimizer. 
\end{abstract}
\maketitle

\section{Introduction}

As a model of an electron moving in an ionic crystal, the polaron
continues to be of interest. Because it is also one of the simplest
examples of a particle interacting with a quantum field, it has served
as a testing ground for various techniques in field theory (see {[}AlDe{]})
such as the Feynman path integral (see {[}Fy{]}). But despite the
attention it has received over the last eight decades, many questions
remain open. With a few exceptions in a limiting case (see e.g. {[}DoVa{]},
{[}LiTh{]}), the polaron has eluded the exact calculation of the most
basic quantitites such as the effective mass and the ground state
energy. Moreover, an exact analytic expression for the ground-state
wave function of the model has yet to be given. In this paper, we
give the first convergence result for the wave function. 

Polaron theory began in the 1930s when ionic crystals of nonmetallic
type- capable of producing strong electric fields- were being introduced
into electronic devices (see {[}Pek{]}). Starting with L.D. Landau's
paper {[}Ld{]} from 1933, where it was suggested that the electron
deforms the crystal and traps itself in a hole of its own making,
various models (see {[}Dev{]}) were developed to explain the transport
of electrons through these crystals. These models were considerably
more reliable for experiments than the standard band theory (see {[}Pek{]}),
because they account for the polarization of the crystal due to the
moving electron. This polarization is modelled in terms of the vibrational
displacement of the ions in the crystal lattice; in the literature
these vibrations are called \textsl{phonon modes}. 

In 1937 H. Fröhlich suggested a model- known today as the \textsl{Fröhlich
polaron}- to explain electrical breakdown in these crystals {[}Fr{]}.
The Hamiltonian is 

\begin{equation}
H_{\alpha}=\mathbf{p}^{2}+\sum_{\mathbf{k}}a_{\mathbf{k}}^{\dagger}a_{\mathbf{k}}-\left(\frac{4\pi\alpha}{\Gamma}\right)^{\frac{1}{2}}\sum_{\mathbf{k}}\left[\frac{a_{\mathbf{k}}}{\left|\mathbf{k}\right|}e^{i\mathbf{k}\cdot\mathbf{x}}+\frac{a_{\mathbf{k}}^{\dagger}}{\left|\mathbf{k}\right|}e^{-i\mathbf{k}\cdot\mathbf{x}}\right],\label{eq:1}
\end{equation}
where \textbf{$\mathbf{p}=-i\nabla$} (the electron momentum) and
acts on $\mathcal{F}\otimes L^{2}(\mathbb{R}^{3})$, where $\mathcal{F}$
denotes the (symmetric) phonon Fock space. In (\ref{eq:1}) we use
``$x$'' to denote the electronic coordinate, ``$k$'' for the
phonon mode and $\Gamma$ for the volume of the crystal. The creation
and annihlation operators $a_{k}^{\dagger}$ and $a_{k}$ are defined
on $\mathcal{F}$ with the canonical commutator relation $[a_{k},a_{k'}^{\dagger}]=\delta(k-k')$.
The coupling paramter $\alpha>0$ was introduced by Fröhlich {[}Fr-2{]}
in 1954 to describe the interaction between the electron and the phonon
modes. The model is also known as the \textsl{large polaron}, because
the spatial extension of the wave function is larger than the crystal
lattice spacing. Therefore a continuum approximation $\sum_{k}\rightarrow\Gamma(2\pi)^{-3}\int d^{3}k$
for the Hamiltonian in (\ref{eq:1}) is also allowed. 

A proof of the self-adjointness for Hamiltonians of this type was
first given in 1964 by E. Nelson {[}Ne{]}. 

The \textsl{ground state energy} of the model is 
\begin{equation}
E_{\alpha}=\inf\left\{ \langle\Psi,\ H_{\alpha}\Psi\rangle_{\mathcal{F}\otimes L^{2}(\mathbb{R}^{3})}\mid\Psi\in\mathcal{F}\otimes L^{2}(\mathbb{R}^{3})\ \mbox{and}\ \|\Psi\|_{\mathcal{F}\otimes L^{2}(\mathbb{R}^{3})}=1\right\} ,\label{eq:2}
\end{equation}
where the operator $H_{\alpha}$ is the Hamiltonian in (\ref{eq:1}).
A normalized wave function in $\mathcal{F}\otimes L^{2}(\mathbb{R}^{3})$
that achieves the ground state energy in (\ref{eq:2}) is a \textsl{ground-state
wave function}. 

The mathematical difficulty in calculating the ground state energy
and the ground-state wave function of the model stems from the electron-phonon
interaction term of the Hamiltonian in (\ref{eq:1}). Not only are
the electron and phonon co-ordinates in (\ref{eq:1}) inseperable,
but we also do not \textit{a priori} know the explicit dependence
between the co-ordinates. The minimization problem in (\ref{eq:2})
is therefore intractable. 

During the 1950s this mathematical difficulty motivated physicists
to develop various techniques for approximating the ground state energy
in (\ref{eq:2}) by exploiting the properties of the ground-state
wave function. In his 1951 monograph {[}Pek{]} S.I. Pekar presents
a non-linear theory posited on his \textsl{Produkt-Ansatz} for the
ground-state wave function. Based on his physical intuition that the
phonons are not sensitive to the instantaneous position of the electron,
Pekar proposed that the ground-state wave function in (\ref{eq:2})
can be expressed in the product form 

\begin{equation}
\Psi=|\phi\rangle\otimes|\zeta\rangle.\label{eq:3}
\end{equation}
In (\ref{eq:3}) $|\zeta\rangle\in\mathcal{F}$ is a coherent state
defined only on the phonon co-ordinates, and $\phi\in L^{2}(\mathbb{R}^{3})$
is a normalized, electronic wave function. In particular, $\alpha^{-\frac{3}{2}}\phi\left(\frac{x}{\alpha}\right)$
is a minimizer of the non-linear problem:

\begin{equation}
e_{P}=\inf\left\{ \int_{\mathbb{R}^{3}}|\nabla\phi|^{2}d\mathbf{x}-\int\int_{\mathbb{R}^{3}\times\mathbb{R}^{3}}\frac{\phi(\mathbf{x})^{2}\phi(\mathbf{y})^{2}}{\left|\mathbf{x}-\mathbf{y}\right|}d\mathbf{x}\, d\mathbf{y}\mid\int_{\mathbb{R}^{3}}\phi(\mathbf{x})^{2}d\mathbf{x}=1\right\} .\label{eq:7}
\end{equation}
Pekar's ansatz in (\ref{eq:3}) offers the computational advantage
of eliminating all of the phonon co-ordinates in the optimization
problem from (\ref{eq:2}) for the ground state energy. Minimizing
$\langle\Psi,\ H_{\alpha}\Psi\rangle$ over the more restrictive set
of product wave functions in (\ref{eq:3}) yields the following \textsl{upper
bound} for the true ground state energy:
\begin{equation}
E_{\alpha}\leq\inf\left\{ \langle\Psi,\ H_{\alpha}(V)\Psi\rangle_{\mathcal{F}\otimes L^{2}(\mathbb{R}^{3})}\mid\|\Psi\|_{\mathcal{F}\otimes L^{2}(\mathbb{R}^{3})}=1\ \mbox{and}\ \Psi=|\phi\rangle\otimes|\zeta\rangle\right\} \label{eq:5}
\end{equation}
\begin{equation}
=\alpha^{2}e_{P},\label{eq:6}
\end{equation}
with the \textsl{Pekar energy} $e_{P}$ as defined in (\ref{eq:7}).
Numerical work in 1976 by S.J. Miyake suggests that $e_{P}=-0.108513$
{[}My{]}.

Pekar's ansatz gives rise to a widely studied minimization problem
with a non-linear energy functional, given in (\ref{eq:7}) above,
known as the \textit{Pekar functional} (see eg. {[}GrHtWl{]}, {[}Li{]},
{[}Ln{]}, {[}Ln-2{]} and {[}LwRg{]}). That a minimizer actually exists
for the non-linear problem in (\ref{eq:7}) has been shown only in
1977 by E.H. Lieb using rearrangment inequalities {[}Li{]}. Lieb has
also established that this minimizer is unique up to translations
by proving that there is a unique solution for the corresponding Euler-Lagrange
equation: 
\begin{equation}
\left\{ -\triangle-2\int_{\mathbb{R}^{3}}|\phi(y)|^{2}|x-y|^{-1}dy\right\} \phi(x)=\lambda\phi(x).\label{eq:200}
\end{equation}

The equation in (\ref{eq:200}) is known as the \textsl{Choquard-Pekar
}or the \textit{Schrödinger-Newton} equation, and it has attracted
a lot of attention in the recent literature (see eg. {[}GmVs{]}, {[}
LwRg{]}, {[}Lz{]}, {[}MzVs{]}, {[}MzVs-2{]}, {[}MzVs-3{]}, {[}MzVs-4{]},
{[}MzVs-5{]} and {[}Rd{]}). In particular, E. Lenzmann has shown that
around the unique minimizer $Q(x)$ of the Choquard-Pekar equation
in (\ref{eq:200}), the linearization 
\[
L_{+}\zeta=-\triangle\zeta+\lambda\zeta-\left(|x|^{-1}\ast|Q|^{2}\right)\zeta-2Q\left(|x|^{-1}\ast(Q\zeta)\right)
\]
has a non-degenerate kernel {[}Lz{]}: 
\begin{equation}
\ker L_{+}=\mbox{span}\{\partial_{x_{1}}Q,\,\partial_{x_{2}}Q,\,\partial_{x_{3}}Q\}.\label{eq:201-1}
\end{equation}
With this non-degeneracy result, Lenzmann establishes the uniqueness
up to translations of the \textit{pseudo-relativistic version} of
the Choquard-Pekar equation via an implicit function-type argument
{[}Lz{]}. Lenzmann's non-degeneracy result in (\ref{eq:201-1}) has
found many applications (see {[}Rd{]} and the references therein),
and it is a crucial ingredient, for example, in the proof of the symmetry
of the bipolaron bound state given by Frank, Lieb and Seiringer in
{[}FrLiSr{]}. These uniqueness and non-degeneracy results have been
extended to the anisotropic polaron (see {[}LwRg{]}) by J. Ricaud
in recent work {[}Rd{]}. 

Despite the mathematical interest that the Pekar functional continues
to stimulate, it was noticed already in the 1950s that Pekar's ansatz
is only physically sensible for describing crystals with a very large
(``strong''-) coupling paramter $\alpha$ {[}Fr-2{]}. And a more
adequate theory was needed to explain the transport of electrons in
weak-coupling crystals (eg. InSb; cf. {[}Dev{]}). In 1954 Fröhlich
has introduced a weak-coupling theory (see {[}Fr-2{]}) based on the
canonical transformation of T.D. Lee, F.E. Low and D. Pines from 1953
{[}LLP{]}. This in turn motivated R.P. Feynman in 1955 to use the
path integral to develop an intermediate-coupling theory that is applicable
to a wider range of coupling paramters {[}Fy{]}. 

These theories- each useful at different values of the coupling parameter-
only provide an upper bound for the true ground state energy $E_{\alpha}$
in (\ref{eq:2}). In 1958, however, E.H. Lieb and K. Yamazaki arrive
at a rigorous lower bound for the ground state energy via modifying
the Hamiltonian (\ref{eq:1}) rather than making an ansatz about the
ground-state wave function {[}LiYa{]}. Their lower bound, however,
differs from Pekar's upper bound in (\ref{eq:6}) by a factor of $3$,
so their calculation unfortunately does not yield the exact ground
state energy in the strong-coupling limit. Lieb and Yamazaki's techniques
from 1958 nevertheless continue to inspire rigorous calculations in
strong-coupling theory: commutator estimates with their vector operator
$\mathbf{Z}=(Z_{1},Z_{2},Z_{3})$, where 
\[
Z_{j}=\left(\frac{4\pi\alpha}{\Gamma}\right)^{\frac{1}{2}}\sum_{|k|>K}k_{j}\frac{a_{k}}{|\mathbf{k}|^{3}}e^{i\mathbf{k}\cdot\mathbf{x}},\ j=1,2,3\ ,
\]
are necessary when using an ultraviolet cutoff on Hamiltonians such
as that in (\ref{eq:1}) (see eg. {[}AnLa{]}, {[}BeBl{]}, {[}FrGs{]},
{[}FrSl{]} and {[}LiTh{]}). It is also useful in quantum electrodynamics
(see proof of Corollary 2.2 in {[}LiLo-2{]}). 

To this day it is not known how to calculate the exact ground state
energy at finite values of the coupling parameter $\alpha$, but in
1981 M.D. Donsker and S.R.S. Varadhan have succeeded in using techniques
from large deviation theory to show that Pekar's approximation ((\ref{eq:5}),
(\ref{eq:6})) of the ground state energy is exact in the strong-coupling
limit {[}DoVa{]}: 
\begin{equation}
\lim_{\alpha\rightarrow\infty}\frac{E_{\alpha}}{\alpha^{2}}=e_{P}.\label{eq:8}
\end{equation}
Of course, $e_{P}$ is the upper bound from (\ref{eq:7}) that is
calculated using Pekar's ansatz for the ground-state wave function.
In 1995 E.H. Lieb and L.E. Thomas provide an alternate, pedestrian
proof of (\ref{eq:8}) using simple modifications of the Hamiltonian,
a philosophy that can be traced back to the earlier work {[}LiYa{]}
of Lieb and Yamazaki from 1958. In {[}LiTh{]} Lieb and Thomas use
coherent states to obtain an agreeable lower bound, and the strategy
also yields a rate of convergence for the result in (\ref{eq:8}).
Their technique has motivated recent study of the ground state energy
of multi-polaron systems (see {[}AnLa{]},{[}BeBl{]} and {[}GrMl{]})
and can also be adapted to other models (see {[}FrGs{]} for the polaron
in a large magnetic field). We use their methods to argue the exact
ground state energy of our one-dimensional model (see Theorem 1 below).

\subsection{Motivation}

Showing Pekar's product wave function yields the exact ground state
energy in the strong-coupling limit is a successful chapter in polaron
theory, spanning more than four decades. This by no means is a justification
of Pekar's ansatz for the ground-state wave function, which, as Lieb
and Yamazaki point out in 1958, is in fact inadequate for calculating
the expectation values of various operators at the ground state. For
example, the expectation value $\langle H_{\alpha}^{2}\rangle=\infty$
when one takes the ground-state wave function to be Pekar's product
function in (\ref{eq:3}). It still remains to understand the connection
between the true ground-state wave function and Pekar's product wave
function. 

With this paper we present a strategy that can be used to show that
in the limit $\alpha\rightarrow\infty$ the true ground-state wave
function of the Fröhlich Hamiltonian- after integrating out its phonon
modes- converges (in the weak sense) to the electronic wave function
in Pekar's ansatz. As evident from (\ref{eq:7}), this electronic
wave function is the minimizer of the corresponding non-linear Pekar
functional. 

A much stronger relationship between the ground state of the Fröhlich
Hamiltonian and the product wave function from Pekar's ansatz in (\ref{eq:3})
has been recently conjectured by E.H. Lieb and R. Seiringer {[}LiSr{]}. 

To have a well-defined notion of convergence, however, we need to
ensure that the corresponding Pekar functional admits a unique minimizer.
Note that the Pekar energy functional in (\ref{eq:7}), originally
studied by Lieb in {[}Li{]}, has translational invariance. To even
have a chance at uniqueness we must break translation invariance by
introducing a localizing electric potential (in the \textbf{x }co-ordinate
only) for the system: for some $V(\mathbf{x})\geq0$ that vanishes
at infinity, consider the localized Hamiltonian 
\begin{equation}
H_{\alpha}(V)\equiv H_{\alpha}-\alpha^{2}V(\alpha\mathbf{x}),\label{eq:100-1}
\end{equation}
where $H_{\alpha}$ is the translation invariant Hamiltonian from
(\ref{eq:1}). The corresponding Pekar energy functional is 
\begin{equation}
\mathcal{E}_{V}(\phi)=\int_{\mathbb{R}^{3}}\left(\left|\nabla\phi\right|^{2}-V(\mathbf{x})\left|\phi(\mathbf{x})\right|^{2}\right)d\mathbf{x}-\int\int_{\mathbb{R}^{3}\times\mathbb{R}^{3}}\frac{\left|\phi(\mathbf{x})\right|^{2}\left|\phi(\mathbf{y})\right|^{2}}{\left|\mathbf{x}-\mathbf{y}\right|}d\mathbf{x}\, d\mathbf{y}.\label{eq:100-3}
\end{equation}

In 1977 E.H. Lieb proved uniqueness \textit{up to translations} {[}Li{]}
for the functional in (\ref{eq:7}) (see also ``Appendix A'' in
{[}Lz{]} for a somewhat different proof), but his proof exploits some
essential scaling relations that are no longer true for the functional
in (\ref{eq:100-3}), where we break translational invariance of the
system with a localizing potential. The existence of a minimizer for
the Pekar functional with the inclusion of an external potential,
(\ref{eq:100-3}), has been studied by P.L. Lions in the 1980s when
developing his compactness arguments (see {[}Ln{]} and {[}Ln-2{]}
and cf. {[}Li-2{]}, an earlier similar result of E.H. Lieb). In the
presence of a localizing potential, while existence issues can now
be settled using Lions' concentration compactness lemma {[}Ln{]},
we are not aware of any tools for addressing the uniqueness of the
minimizer. Even when the localizing potential in (\ref{eq:100-3})
is the Coulomb potential, $|x|^{-1}$, we are unable to adapt the
technique from {[}Li{]}. 

A uniqueness result is available only in one-dimension (see {[}JjSt{]}
and {[}MlStTr{]}), so we can only provide rigorous arguments for a
one-dimensional model. But we emphasize that there is nothing intrinsically
one-dimensional about our strategy, which follows from a very simple
application of the variational principle.

\section{The One-Dimensional Model and Statement of Results}

We work with the one-dimensional Fröhlich polaron localized in a \textit{symmetric
decreasing}, $C^{1}(\mathbb{R})$ electric potential $V(x)\geq0$
that \textit{vanishes at infinity}. The Hamiltonian for our model,

\begin{equation}
H_{\alpha}(V)=-\frac{d^{2}}{dx^{2}}+\sum_{|k|>0}a_{k}^{\dagger}a_{k}-\left(\frac{\alpha}{L}\right)^{\frac{1}{2}}\sum_{|k|>0}\left[a_{k}e^{ikx}+a_{k}^{\dagger}e^{-ikx}\right]-\alpha^{2}V(\alpha x),\label{eq:9999}
\end{equation}
is defined on $\mathcal{F}\otimes L^{2}(\mathbb{R})$, where $\mathcal{F}$
is a symmetric Fock space over $\ell^{2}\left(\mathbb{Z}/L\right)$.
In (\ref{eq:9999}) we use ``$x$'' to denote the electronic coordinate,
``$k$'' for the phonon mode, ``$2L$'' is the length of the crystal
and ``$\frac{1}{L}$'' is the lattice spacing. The creation and
annihlation operators $a_{k}^{\dagger}$ and $a_{k}$ are defined
on $\mathcal{F}$ with the canonical commutator relation $[a_{k},a_{k'}^{\dagger}]=\delta(k-k')$.
As usual the continuum approximation ``$\sum_{k}\rightarrow L\int dk$''
is allowed. 

The ground state energy of the model is defined as 
\begin{equation}
E_{\alpha}(V)=\inf\{\langle\Psi,\ H_{\alpha}(V)\Psi\rangle_{\mathcal{F}\otimes L^{2}}\mid\|\Psi\|_{\mathcal{F}\otimes L^{2}}=1\},\label{eq:10}
\end{equation}
and an optimizing function in (\ref{eq:10}) is the ground state wave
function of the model. As with the three-dimensional case discussed
above, it is straightforward to calculate with Pekar's \textit{Produkt-Ansatz}
(see (\ref{eq:3}) above) for the ground-state wave function
\begin{equation}
\Psi=|u\rangle_{L^{2}(\mathbb{R})}\otimes|\zeta\rangle_{\mathcal{F}},\label{eq:ansatz}
\end{equation}
 that 
\begin{equation}
E_{\alpha}(V)\leq\alpha^{2}e(V).\label{eq:11}
\end{equation}
A scaled version of the electronic wave function in Pekar's ansatz,
$\alpha^{-\frac{1}{2}}u(\frac{x}{\alpha})$, is a minimizer of the
one-dimensional minimization problem for the Pekar energy $e(V)$: 

\begin{equation}
e(V)=\inf\{\mathcal{E}_{V}(u):\ \int_{\mathbb{R}}u^{2}dx=1\},\label{eq:12}
\end{equation}
with the one-dimensional Pekar functional: 
\begin{equation}
\mathcal{E}_{V}(u)=\int_{\mathbb{R}}\left(u'^{2}-u^{4}-V(x)u^{2}\right)dx.\label{eq:0}
\end{equation}
When $V=0,$ it has been shown that $e(0)=-\frac{1}{12}$ (see e.g.
{[}Gh{]}). Since $V$ is a localizing potential, $e(V)<0$. 

The one-dimensional polaron was introduced by E.P. Gross in his 1976
paper as a toy model {[}Go{]}. But it has since attracted sizeable
attention in the literature and is not entirely artificial (see eg.
{[}FrGs{]}, {[}FrLiSrTh{]}, {[}Gn{]}, {[}KoLeSm{]}, {[}Mn{]}, {[}PtSm{]},
{[}SKVPD{]}, {[}Sp{]}, {[}Sp-2{]}, {[}VPSD{]}). The most compelling
reason to study the one-dimensional model is its connection to the
three-dimensional polaron in a magnetic field, also known as the \textsl{magneto-polaron}.
In 1992 E.A. Kochetov, H. Leschke and M.A. Smondyrev {[}KoLeSm{]}
consider a three-dimensional polaron in a strong magnetic field $\mathbf{B}$
and with some fixed coupling $\alpha>0$. In the limit $|\mathbf{B}|\rightarrow\infty$,
they argue that the dynamics of the three-dimensional magneto-polaron
is equivalent to that of a one-dimensional strong-coupling polaron
with the large coupling constant $\alpha'=\left(\alpha\ln|\mathbf{B}|\right)/2$. 

But Kochetov et. al's argument- reminiscent of Pekar's heuristic justification
{[}Pek{]} of his \textit{Produkt-Ansatz}- has been rigorously verified
only recently by R.L. Frank and L. Geisinger in {[}FrGs{]}. They calculate
the exact ground state energy of the magneto-polaron and use the one-dimensional
version of the strategy developed in {[}LiTh{]}. We present this one-dimensional
version in full detail in order to calculate the exact ground state
energy of our model. We also think this calculation conveniently complements
Frank and Geisinger's argument in Section 6 of their paper {[}FrGs{]}. 
\begin{thm}
For the ground state energy $E_{\alpha}(V)$ in (\ref{eq:10}) and
the Pekar energy e(V) in (\ref{eq:12}), 
\[
\lim_{\alpha\rightarrow\infty}\frac{E_{\alpha}(V)}{\alpha^{2}}=e(V).
\]

\end{thm}
The first natural task is to see that a true ground-state wave function
exists. While there are existence results in the literature (see {[}Sp{]}
{[}Sp-2{]}, {[}Sp-3{]}, {[}GrLo{]} and {[}GrLo-2{]}), we circumvent
this important issue by using our result in Theorem 1 to define the
notion of an \textit{approximate ground-state wave function}. 
\begin{defn}
A wave function $\Psi_{\alpha}\in\mathcal{F}\otimes L^{2}(\mathbb{R})$
is an \textit{approximate ground-state wave function} for the localized
Hamiltonian $H_{\alpha}(V)$ in (\ref{eq:9999}) if $\|\Psi_{\alpha}\|_{\mathcal{F\otimes}L^{2}(\mathbb{R})}=1$
and 
\begin{equation}
\langle\Psi_{\alpha},\, H_{\alpha}(V)\Psi_{\alpha}\rangle_{\mathcal{F}\otimes L^{2}(\mathbb{R})}-E_{\alpha}(V)=o(\alpha^{2}).\label{eq:20}
\end{equation}

\end{defn}
If a true ground-state wave function exists, it is obviously also
an approximate ground-state wave function. From Theorem 1 we see that
Pekar's product wave function in (\ref{eq:ansatz}) is also an approximate
ground-state wave function. We will show that \textit{any} approximate
ground-state wave function of our model converges in the weak sense
to the \textit{unique} minimizer of the corresponding Pekar functional
given in (\ref{eq:0}). 

Because the potential $V(x)$ is symmetric decreasing and vanishes
at infinity, it can easily be shown using the direct method in the
calculus of variations that a minimizer exists for the minimization
problem in (\ref{eq:12}) (see Theorem 8.6 and Chapter 11 in {[}LiLo{]};
see also {[}Gh{]}). The argument is very similar to our proof of Lemma
5 below. A minimizer is also real-valued and positive (cf. Theorem
7.8 in {[}LiLo{]}). 

To argue uniqueness we consider the Euler-Lagrange equation. Any minimizer
of the problem in (\ref{eq:12}) is a positive, real-valued solution
of the eigenvalue equation 

\begin{equation}
-u''-2u^{3}-Vu=\lambda u,\label{eq:29-1}
\end{equation}
where $\lambda\in\mathbb{R}$ is a Lagrange multiplier corresponding
to the constraint $\int_{\mathbb{R}}u^{2}dx=1$ in (\ref{eq:12}).
It can be argued with the standard tools in regularity theory (Theorem
11.7 in {[}LiLo{]}; see also {[}Gh{]}) that the solution of (\ref{eq:29-1})
is in fact a classical solution. If there are two minimizers for the
problem in (\ref{eq:12}), then they each satisfy (\ref{eq:29-1})
with possibly different values of $\lambda$. To show that our one-dimensional
Pekar functional from (\ref{eq:12}) admits a unique minimizer, we
must see that over all non-negative $u\in L^{2}(\mathbb{R})$ with
$\int_{\mathbb{R}}u^{2}dx=1$ there is only one pair $(\lambda,u)$
which satisfies the equation in (\ref{eq:29-1}). We see this from
Theorem 3 in {[}JjSt{]} and Theorem 2.1 in {[}MlStTr{]}:
\begin{prop}
If $V\in C^{1}(\mathbb{R})$ is a nonzero, nonnegative and symmetric
decreasing function that vanishes at infinity, then the one-dimensional
minimization problem in (\ref{eq:12}) for the Pekar energy $e(V)$
admits a unique minimizer. 
\end{prop}
We briefly describe their proofs. First, consider the lowest eigenvalue
of the linearization of the equation in (\ref{eq:29-1}): 

\begin{equation}
\lambda_{0}=\inf\left\{ \int_{-\infty}^{+\infty}(u')^{2}-V(x)u^{2}dx:\ u\in H^{2}(\mathbb{R})\ \mbox{and}\ \int_{-\infty}^{\infty}u^{2}dx=1\right\} .\label{eq:31-2}
\end{equation}
In 1999, H. Jeanjean and C.A. Stuart bifurcate (see Theorem 3 in {[}JjSt{]})
a unique continuous curve of solutions $u\in C^{1}((-\infty,\lambda_{0}),\lambda)$
such that for each $\lambda\in(-\infty,\lambda_{0})$, the pair $(\lambda,u(\lambda))$
is a solution of (\ref{eq:29-1}). They argue that all solutions of
(\ref{eq:29-1}) belong to this curve: 

\[
\{(\lambda,u(\lambda)):\ \lambda\in(-\infty,\lambda_{0})\}=\{(\mu,\nu):\ (\mu,\nu)\ \mbox{is a solution to }(\ref{eq:29-1})\}.
\]
This means that for each $-\infty<\lambda<\lambda_{0}$, there is
a unique $u(\lambda)$ that satisfies (\ref{eq:29-1}). In 2003 J.B.
McLeod, C.A. Stuart and W.C. Troy show (Theorem 2.1 in {[}MlStTr{]})
that $\|u(\lambda)\|_{L^{2}(\mathbb{R})}$ decreases as $\lambda$
increases from $-\infty$ to $\lambda_{0}$. Therefore, there is only
one pair $(\lambda,u(\lambda))$ that satisfies (\ref{eq:29-1}) with
$\|u(\lambda)\|_{2}=1$, and this is the unique minimizer of our problem
in (\ref{eq:12}). 

After bifurcating a curve of solutions from the lowest eigenvalue
$\lambda_{0}$ in (\ref{eq:31-2}), the strategy in {[}JjSt{]} is
to repeatedly use the implicit function theorem at the positive solutions
to get a global branch containing all the positive, real-valued solutions
of the equation in (\ref{eq:29-1}). As it was pointed out to the
author by Professor C.A. Stuart, the uniqueness result is limited
to one-dimension, because the authors in {[}JjSt{]} can only see how
to check the hypotheses of the implicit function theorem in the one-dimensional
case. Local bifurcation at the lowest eigenvalue is possible in all
dimensions since the eigenvalue is always simple (cf. {[}CrRa{]}),
but global continuation is the main problem. 

With this uniqueness result we have a well-defined notion of convergence,
and we now state our main result. 

(In this paper, we say a Borel measure $W$ is \textit{bounded} if
$\int_{\mathbb{R}}W(x)dx=1$. An example is the Dirac-delta measure.)
\begin{thm}
Let $u_{V}$ be the unique minimizer of the minimization problem in
(\ref{eq:12}), and let $\Psi_{\alpha}\in\mathcal{F}\otimes L^{2}(\mathbb{R})$
be any approximate ground-state wave function of the Hamiltonian in
(\ref{eq:9999}). Then: 

\begin{equation}
\lim_{\alpha\rightarrow\infty}\int_{\mathbb{R}}\left[\frac{1}{\alpha}\|\Psi_{\alpha}\left(\frac{\cdot}{\alpha}\right)\|_{\mathcal{F}}^{2}\right]W(x)dx=\int_{\mathbb{R}}|u_{V}|^{2}W(x)dx\label{eq:m1}
\end{equation}
for any bounded Borel measure $W(x)$ on the real line. 
\end{thm}

\subsection{Strategy}

Let $\delta$ be a real parameter and $W(x)$ some bounded Borel measure
as above. The main idea, suggested to the author by Professor R.L.
Frank, is to perturb the Hamiltonian $H_{\alpha}(V)$ in (\ref{eq:9999}):

\begin{equation}
H_{\alpha}(V+\delta W)\equiv H_{\alpha}(V)+\alpha^{2}\delta W(\alpha x).\label{eq:s1}
\end{equation}
Imitating the proof of Theorem 1, we calculate the exact ground state
energy of the ``perturbed Hamiltonian'' in (\ref{eq:s1}):
\begin{equation}
E_{\alpha}(V+\delta W)=\alpha^{2}e(V+\delta W),\label{eq:s2}
\end{equation}

\begin{equation}
e(V+\delta W)\equiv\inf\left\{ \mathcal{E}_{V+\delta W}(u):\ \int_{\mathbb{R}}u^{2}dx=1\right\} ,\label{eq:s3}
\end{equation}
where 
\begin{equation}
\mathcal{E}_{V+\delta W}(u)\equiv\mathcal{E}_{V}(u)+\delta\int_{\mathbb{R}}W(x)|u|^{2}dx.\label{eq:s4}
\end{equation}
Of course the Pekar functional $\mathcal{E}_{V}$ was already defined
in (\ref{eq:12}), and from Proposition 3 we know it admits a unique
minimizer $u_{V}$. 

Let $\Psi_{\alpha}$ be an approximate ground-state wave function
of the Hamiltonian $H_{\alpha}(V)$. Since $\Psi_{\alpha}$ is not
necessarily the ground-state wave function of the perturbed Hamiltonian
$H_{\alpha}(V+\delta W)$, a simple application of the variational
principle yields
\[
E_{\alpha}(V+\delta W)\leq\langle\Psi_{\alpha},\,\left[H_{\alpha}(V+\delta W)\right]\Psi_{\alpha}\rangle
\]
\[
=\langle\Psi_{\alpha},H_{\alpha}(V)\Psi_{\alpha}\rangle+\alpha^{2}\delta\langle\Psi_{\alpha},W(\alpha x)\Psi_{\alpha}\rangle
\]
\[
\leq E_{\alpha}(V)+o(\alpha^{2})+\alpha^{2}\delta\langle\Psi_{\alpha},W(\alpha x)\Psi_{\alpha}\rangle,
\]
where the inner product is on $\mathcal{F}\otimes L^{2}(\mathbb{R})$. 

Dividing by $\alpha^{2}$ and using Theorem 1, (\ref{eq:s2}) and
(\ref{eq:s3}) we see that 
\begin{equation}
\frac{e(V+\delta W)-e(V)}{\delta}=\lim_{\alpha\rightarrow\infty}\int_{\mathbb{R}}\left[\frac{1}{\alpha}\|\Psi_{\alpha}\left(\frac{\cdot}{\alpha}\right)\|_{\mathcal{F}}^{2}\right]W(x)\, dx.\label{eq:s66}
\end{equation}
Then our main result (\ref{eq:m1}) in Theorem 4 above follows directly
from (\ref{eq:s66}) if we can differentiate the ``perturbed Pekar
energy'' $e(V+\delta W)$ at $\delta=0$: 
\begin{lem}
Suppose the one-dimensional minimization problem in (\ref{eq:12})
for the Pekar energy $e(V)$ admits a unique minimizer $u_{V}$. Let
$W(x)$ be any bounded Borel measure on the real line, and let $e(V+\delta W)$
be the perturbed Pekar energy defined in (\ref{eq:s3}) with some
real parameter $\delta$. Then the map $\delta\mapsto e(V+\delta W)$
is differentiable at $\delta=0$ and 

\[
\left.\frac{d}{d\delta}\right|_{\delta=0}e(V+\delta W)=\int_{\mathbb{R}}W(x)|u_{V}|^{2}dx.
\]

\end{lem}
The proof of Lemma 5- which we provide in Section 4 below- again follows
from a simple application of the variational principle and a compactness
argument (Theorem 8.6 in {[}LiLo{]}).

\subsection*{Acknowledgement}

I thank Professor Rupert L. Frank for suggesting this problem, for
informing me about the crucial uniqueness result from {[}JjSt{]} and
for advising me during my undergraduate studies. I thank Professor
Michael Loss for his encouragement and many helpful discussions.

\section{Exact Ground State Energy: Proof of Theorem 1}

In this section we will calculate the exact ground state energy of
our one-dimensional model using the strategy developed in {[}LiTh{]}. 

We use $\langle H\rangle$ to denote the expectation value of the
operator on $\mathcal{F}\otimes L^{2}(\mathbb{R})$. Likewise, $\|\Psi\|$
should be read: $\|\Psi\|_{\mathcal{F}\otimes L^{2}(\mathbb{R})}$. 

The main idea in this proof is to work with a rigorously justified
approximation that the electron only interacts with finitely many
phonon modes and to then use coherent states to arrive at a lower
bound that agrees- to the leading order in $\alpha$- with Pekar's
upper bound in (\ref{eq:11}). This already departs from the physical
picture offered by Pekar's ansatz, that the phonons cannot follow
the electron. 

The general utility of coherent states for obtaining rigorous (lower)
bounds is discussed in {[}LiSrYg{]} and {[}Li-3{]}.

\subsection{Ultraviolet Cutoff. }

We ignore large modes in the Fröhlich Hamiltonian and work on a reduced
mode space $\{k:\ |k|\leq K\}$ with a cutoff Hamiltonian 
\begin{equation}
H_{K}=(1-\epsilon)p^{2}+\sum_{|k|\leq K}a_{k}^{\dagger}a_{k}-\left(\frac{\alpha}{L}\right)^{\frac{1}{2}}\sum_{|k|\leq K}\left(a_{k}e^{ikx}+a_{k}^{\dagger}e^{-ikx}\right)-\alpha^{2}V(\alpha x)\label{eq:1-1}
\end{equation}

The parameters $\epsilon$ and $K$ will be chosen at the very end
of the computations. 

We bound the energy of the Fröhlich Hamiltonian from below with the
energy of the \textit{cutoff Hamiltonian} $H_{K}$ in (\ref{eq:1-1}).
We observe
\begin{equation}
H_{\alpha}(V)=H_{K}+\epsilon p^{2}+\sum_{|k|>K}a_{k}^{\dagger}a_{k}-\left(\frac{\alpha}{L}\right)^{\frac{1}{2}}\sum_{|k|>K}\left(a_{k}e^{ikx}+a_{k}^{\dagger}e^{-ikx}\right)\label{eq:2-11}
\end{equation}
and we will arrive at a lower bound by making an estimate on the interaction
term in (\ref{eq:2-11}). 

We use a standard commutator identity $\langle[p,a_{k}e^{ikx}]\rangle=k\langle a_{k}e^{ikx}\rangle$
and work with two operators $Z=\left(\frac{\alpha}{L}\right)^{\frac{1}{2}}\underset{|k|>K}{\sum}\frac{a_{k}e^{ikx}}{k}$
and $Z^{\dagger}=\left(\frac{\alpha}{L}\right)^{\frac{1}{2}}\underset{|k|>K}{\sum}\frac{a_{k}^{\dagger}e^{-ikx}}{k}$. 

For any $0<\epsilon<1$, 
\[
\left\langle \left(\frac{\alpha}{L}\right)^{\frac{1}{2}}\sum_{|k|>K}\left(a_{k}e^{ikx}+a_{k}^{\dagger}e^{-ikx}\right)\right\rangle =\left\langle [p,Z-Z^{\dagger}]\right\rangle 
\]
\[
\leq2\left\langle p^{2}\right\rangle ^{\frac{1}{2}}\left\langle -(Z-Z^{\dagger})^{2}\right\rangle ^{\frac{1}{2}}
\]
\[
\leq\epsilon\left\langle p^{2}\right\rangle +\frac{1}{\epsilon}\left\langle -(Z-Z^{\dagger})^{2}\right\rangle 
\]
\begin{equation}
\leq\epsilon\left\langle p^{2}\right\rangle +\frac{2}{\epsilon}\left\langle ZZ^{\dagger}+Z^{\dagger}Z\right\rangle \label{eq:3-11}
\end{equation}
\[
=\epsilon\left\langle p^{2}\right\rangle +\frac{2}{\epsilon}\left\langle \left[Z,Z^{\dagger}\right]\right\rangle +\frac{4}{\epsilon}\left\langle Z^{\dagger}Z\right\rangle 
\]
\begin{equation}
\leq\epsilon\left\langle p^{2}\right\rangle +\frac{4\alpha}{\epsilon K}+\frac{8\alpha}{\epsilon K}\left\langle \sum_{|k|>K}a_{k}^{\dagger}a_{k}\right\rangle \label{eq:4}
\end{equation}

Above, (\ref{eq:3-11}) is immediate from the positive definiteness
of $(Z+Z^{\dagger})^{2}.$ To arrive at (\ref{eq:4}), we make the
following estimates on $\left\langle Z^{\dagger}Z\right\rangle $
and $\left\langle [Z,Z^{\dagger}]\right\rangle $: 
\[
\left\langle Z^{\dagger}Z\right\rangle =\left(\frac{\alpha}{L}\right)\sum_{|k|>K}\sum_{|k'|>K}\left\langle \frac{a_{k}^{\dagger}a_{k'}e^{i(k'-k)x}}{kk'}\right\rangle 
\]
\[
\leq\left(\frac{\alpha}{L}\right)\left(\sum_{|k|>K}\frac{1}{k}\langle a_{k}^{\dagger}a_{k}\rangle^{\frac{1}{2}}\right)^{2}
\]
\[
\leq\left(\frac{\alpha}{L}\right)\left(\sum_{|k|>K}\frac{1}{|k|^{2}}\right)\left(\sum_{|k|>K}\langle a_{k}^{\dagger}a_{k}\rangle\right)
\]
\[
\leq\frac{2\alpha}{K}\left\langle \sum_{|k|>K}a_{k}^{\dagger}a_{k}\right\rangle 
\]
Since $[a_{k},a_{k'}^{\dagger}]=\delta_{kk'}$, 
\[
\left\langle \left[Z,Z^{\dagger}\right]\right\rangle =\left(\frac{\alpha}{L}\right)\sum_{|k|>K}\sum_{|k'|>K}\left\langle \frac{e^{i(k-k')x}(a_{k}a_{k'}^{\dagger}-a_{k'}^{\dagger}a_{k})}{kk'}\right\rangle 
\]
\[
\leq\left\langle \frac{2\alpha}{K}\right\rangle 
\]

We can now construct a lower bound from (\ref{eq:2-11}): 
\[
\langle H_{\alpha}(V)\rangle\geq\left\langle H_{K}\right\rangle +\left(1-\frac{8\alpha}{\epsilon K}\right)\left\langle \sum_{|k|>K}a_{k}^{\dagger}a_{k}\right\rangle -\left\langle \frac{4\alpha}{\epsilon K}\right\rangle 
\]
Clearly, we require from our parameters $\epsilon$ and $K$ that
$1=\frac{8\alpha}{\epsilon K}$. We now arrive at our lower bound:
\[
E_{\alpha}(V)\geq\inf_{\|\Psi\|=1}\left\langle \Psi,\ H_{K}\Psi\right\rangle -\frac{1}{2}.
\]
In sharp contrast to the three-dimensional computation performed in
{[}LiTh{]}, our error term, $-\frac{1}{2}$, does not depend on the
cutoff parameter $K$.

\subsection{Localizing the Electron. }

We will bound\textit{ from below} the ground state energy of the cutoff
Hamiltonian $H_{K}$: $\underset{\|\Psi\|=1}{\inf}\left\langle \Psi,\ H_{k}\Psi\right\rangle $
(given in (\ref{eq:1-1})). 

Here, $(\triangle E)>0$ is a parameter whose specific value will
be given at the very end of the computations. We will denote by $\underset{\|\Psi\|}{\inf\,'}\left\langle \Psi,\ H_{K}\Psi\right\rangle $
the infimum taken over all wave functions whose \textit{electronic
coordinate is localized} in an interval of length $\frac{\pi}{(\triangle E)^{\frac{1}{2}}}$.
This restriction, we argue, increases the ground state energy of $H_{K}$
by atmost $\triangle E$: 

\begin{equation}
\inf_{\|\Psi\|=1}\left\langle \Psi,\ H_{K}\Psi\right\rangle \geq\underset{\|\Psi\|=1}{\inf\,'}\left\langle \Psi,\ H_{K}\Psi\right\rangle -(\triangle E)\label{eq:5-11}
\end{equation}

Let $\|\Psi\|=1$ and $E=\left\langle \Psi,\ H_{K}\Psi\right\rangle $.
We define $\phi(x)=\cos\left((\triangle E)^{\frac{1}{2}}x\right)$
on its support in $\left(-\frac{\pi}{2(\triangle E)^{\frac{1}{2}}},\frac{\pi}{2(\triangle E)^{\frac{1}{2}}}\right)$
and write $\phi_{y}(x)=\phi(x-y)$. To argue (\ref{eq:5-11}), it
suffices to show for some $\bar{y}\in\mathbb{R}$, 
\begin{equation}
\frac{\left\langle \left(\phi_{\bar{y}}\Psi\right),\ H_{K}\left(\phi_{\bar{y}}\Psi\right)\right\rangle }{\left\langle \phi_{\bar{y}}\Psi,\ \phi_{\bar{y}}\Psi\right\rangle }\leq E+(\triangle E).\label{eq:6-1}
\end{equation}

A direct calculation gives $\int_{\mathbb{R}}\left\langle \left(\phi_{y}\Psi\right),\ H_{K}\left(\phi_{y}\Psi\right)\right\rangle dy=\int(\phi')^{2}+E\phi^{2}dx$
and 
\[
\int_{\mathbb{R}}\left(\left\langle \left(\phi_{y}\Psi\right),\ H_{K}\left(\phi_{y}\Psi\right)\right\rangle -(E+\triangle E)\left\langle \phi_{y}\Psi,\ \phi_{y}\Psi\right\rangle \right)dy
\]
\[
=\int_{-\frac{\pi}{2(\triangle E)^{\frac{1}{2}}}}^{\frac{\pi}{2(\triangle E)^{\frac{1}{2}}}}(\phi')^{2}-(\triangle E)\phi^{2}dx=0
\]
since $(\triangle E)$ is the Dirichlet energy of $\phi$. So there
exists some $\bar{y}\in\mathbb{R}$ such that (\ref{eq:6-1}) holds.
From now on, we consider the electron to be localized in some interval
of length $\frac{\pi}{(\triangle E)^{\frac{1}{2}}}$.

\subsection{Block Hamiltonian.}

We now decompose our finite mode space into finitely many blocks:
$\{k:\ |k|<K\}=\underset{n}{\bigcup}\{B_{n}\}$; each block $B_{n}$
contains ``$PL$'' modes where ``$\underset{k_{i},k_{j}\in B_{n}}{\max}|k_{i}-k_{j}|=P$''
is the size of each block. 

On each block $B_{n}$ we analogously define block annihlation and
creation operators: $A_{B_{n}}=\frac{1}{(PL)^{\frac{1}{2}}}\underset{k\in B_{n}}{\sum}a_{k}$
and $A_{B_{n}}^{\dagger}=\frac{1}{(PL)^{\frac{1}{2}}}\underset{k\in B_{n}}{\sum}a_{k}^{\dagger}$.
Clearly, $[A_{B_{m}},A_{B_{n}}^{\dagger}]=\delta_{mn}.$ On each block
$B_{n}$, we see from the Cauchy-Schwarz inequality that 
\begin{equation}
A_{B_{n}}^{\dagger}A_{B_{n}}\leq\sum_{k\in B_{n}}a_{k}^{\dagger}a_{k}.\label{eq:9-1}
\end{equation}

For reasons that will become clear in the next stage of the computation,
we now want to work with the approximation that the electron interacts
with atmost one mode $k_{B_{n}}$ in each block $B_{n}$. For this
approximation to work, we make the following estimate on the interaction
term of our cutoff Hamiltonian: for any parameter $0<\delta<1$ and
any mode $k_{B_{n}}$ in each block $B_{n}$, completing the square
yields 
\[
\left\langle \left(\frac{\alpha}{L}\right)^{\frac{1}{2}}\sum_{B_{n}}\sum_{k\in B_{n}}\left[a_{k}\left(e^{ik_{B_{n}}x}-e^{ikx}\right)+a_{k}^{\dagger}\left(e^{-ik_{B_{n}}x}-e^{-ikx}\right)\right]\right\rangle 
\]
\[
\leq\left\langle \delta\sum_{|k|<K}a_{k}^{\dagger}a_{k}+\left(\frac{\alpha}{L}\right)\frac{1}{\delta}\sum_{B_{n}}\sum_{k\in B_{n}}\left|e^{ik_{B_{n}}x}-e^{ikx}\right|^{2}\right\rangle 
\]
\[
\leq\left\langle \delta\sum_{|k|<K}a_{k}^{\dagger}a_{k}+\left(\frac{\alpha}{L}\right)\frac{1}{\delta}\sum_{B_{n}}\sum_{k\in B_{n}}|k-k_{B_{n}}|^{2}|x|^{2}\right\rangle 
\]
\begin{equation}
\leq\left\langle \delta\sum_{|k|<K}a_{k}^{\dagger}a_{k}\right\rangle +\frac{2\alpha KP^{2}\pi^{2}}{\delta(\triangle E)}.\label{eq:7-1}
\end{equation}
To arrive at (\ref{eq:7-1}), we used the rigorously justified approximation
(see (\ref{eq:5-11})) that the electronic co-ordinate is localized
in an interval of length $\frac{\pi}{(\triangle E)^{\frac{1}{2}}}$. 

The parameters $P$ and $0<\delta<1$ will be chosen at the very end
of the computations; the specific mode $k_{B_{n}}$ in each block
will be chosen in the next stage of the computation. 

Now we bound the ground state energy of the cutoff Hamiltonian (with
the condition that the electronic co-ordinate of the ground-state
wave function is localized) $\underset{\|\Psi\|=1}{\inf\,'}\left\langle \Psi,\ H_{K}\Psi\right\rangle $,
from below, using the energy of the \textit{block Hamiltonian}: 
\[
H_{K}^{\mbox{Block}}(\{k_{B_{n}}\})
\]
\begin{equation}
=(1-\epsilon)p^{2}+(1-\delta)\sum_{B_{n}}A_{B_{n}}^{\dagger}A_{B_{n}}-(P\alpha)^{\frac{1}{2}}\sum_{B_{n}}\left(A_{B_{n}}e^{ik_{B_{n}}x}+A_{B_{n}}^{\dagger}e^{-ik_{B_{n}}x}\right)-\alpha^{2}V(\alpha x).\label{eq:bl-1}
\end{equation}

Clearly, 
\[
H_{K}=\left((1-\epsilon)p^{2}+\sum_{|k|<K}a_{k}^{\dagger}a_{k}-\left(\frac{\alpha}{L}\right)^{\frac{1}{2}}\sum_{B_{n}}\sum_{k\in B_{n}}\left(a_{k}e^{ik_{B_{n}}x}+a_{k}^{\dagger}e^{-ik_{B_{n}}x}\right)+\right.
\]
\[
\left.-\alpha^{2}V(\alpha x)+\left(\frac{\alpha}{L}\right)^{\frac{1}{2}}\sum_{B_{n}}\sum_{k\in B_{n}}\left[a_{k}\left(e^{ik_{B_{n}}x}-e^{ikx}\right)+a_{k}^{\dagger}\left(e^{-ik_{B_{n}}x}-e^{-ikx}\right)\right]\right)
\]
\begin{equation}
\geq H_{K}^{\mbox{Block}}\left(\{k_{B_{n}}\}\right)-\frac{2\alpha KP^{2}\pi^{2}}{\delta(\triangle E)},\label{eq:8-1}
\end{equation}
in the sense of expectation values. To arrive at (\ref{eq:8-1}) we
simply used the estimates from (\ref{eq:9-1}) and (\ref{eq:7-1}). 

We summarize: 
\[
\underset{\|\Psi\|=1}{\inf\,'}\left\langle \Psi,\ H_{K}\Psi\right\rangle \geq\underset{\|\Psi\|=1}{\inf\,'}\sup_{\{k_{B_{n}}\}}\left\langle \Psi,\ H_{K}^{\mbox{Block}}\left(\left\{ k_{B_{n}}\right\} \right)\Psi\right\rangle -\frac{2\alpha KP^{2}\pi^{2}}{\delta(\triangle E)}.
\]

\subsection{Coherent States. }

We now work with the block Hamiltonian $H_{K}^{\mbox{Block}}\left(\left\{ k_{B_{n}}\right\} \right)$
from (\ref{eq:bl-1}) and the block creation and annihlation operators
constructed in the previous stage of the computation. For each block
$B_{n}$ we define a block coherent state indexed by some $z_{B_{n}}\in\mathbb{C}$:
\[
\left|z_{B_{n}}\right\rangle =\pi^{-\frac{1}{2}}\left(e^{-\frac{|z_{B_{n}}|^{2}}{2}+z_{B_{n}}A_{B_{n}}^{\dagger}}\right)\left|0_{B_{n}}\right\rangle ,
\]
 where $\left|0_{B_{n}}\right\rangle $ denotes the vacuum state in
block $B_{n}$, i.e., $A_{B_{n}}\left|0_{B_{n}}\right\rangle =0$.
We write 
\[
\left|z\right\rangle =\underset{B_{n}}{\prod}\left|z_{B_{n}}\right\rangle ,
\]
a tensor product of the coherent states corresponding to each block
$B_{n}$. 

One can verify that for each block $B_{n}$, the coherent state $\left|z_{B_{n}}\right\rangle $
is the eigenstate of the corresponding block annihlation operator:
\[
A_{B_{n}}\left|z_{B_{n}}\right\rangle =z_{B_{n}}\left|z_{B_{n}}\right\rangle .
\]
The commutator relation $[A_{B_{m}},A_{B_{n}}^{\dagger}]=\delta_{mn}$,
together with the resolution of identity formula 
\[
I=\int\left|z\right\rangle \left\langle z\right|\prod_{B_{m}}dz_{B_{m}}d\overline{z_{B_{m}}},
\]
yield the convenient representations: 
\[
A_{B_{n}}=\int z_{B_{n}}\left|z\right\rangle \left\langle z\right|\prod_{B_{m}}dz_{B_{m}}d\overline{z_{B_{m}}}
\]
\[
A_{B_{n}}^{\dagger}A_{B_{n}}=\int\left(\left|z_{B_{n}}\right|^{2}-1\right)\left|z\right\rangle \left\langle z\right|\prod_{B_{m}}dz_{B_{m}}d\overline{z_{B_{m}}}
\]

Denoting $\Psi_{z}(x)=\left\langle z\mid\Psi\right\rangle _{\mbox{Phonon}}$(the
inner product only in the phonon coordinates), we recast the energy
of the block Hamiltonian $H_{K}^{\mbox{Block}}\left(\left\{ k_{B_{n}}\right\} \right)$
in the following form: 
\begin{equation}
\left\langle \Psi,\ H_{K}^{\mbox{Block}}\left(\left\{ k_{B_{n}}\right\} \right)\Psi\right\rangle =\int\left\langle \Psi_{z},\ h_{z}\Psi_{z}\right\rangle _{\mbox{E}}\prod_{B_{m}}dz_{B_{m}}d\overline{z_{B_{m}}}\label{eq:11-1}
\end{equation}
where $\left\langle \cdot,\cdot\right\rangle _{\mbox{E}}$ is the
inner product only over the electronic coordinate, and $h_{z}$ is
the Schrödinger operator: 
\begin{equation}
h_{z}=(1-\epsilon)p^{2}+\sum_{B_{n}}\left[(1-\delta)\left(\left|z_{B_{n}}\right|^{2}-1\right)-(P\alpha)^{\frac{1}{2}}\left(z_{B_{n}}e^{ik_{B_{n}}x}+\overline{z_{B_{n}}}e^{-ik_{B_{n}}x}\right)\right]-\alpha^{2}V(\alpha x)\label{eq:10-1}
\end{equation}

Since 
\[
\left(\langle\Psi_{z},\Psi_{z}\rangle_{E}(1-\delta)\overline{z_{B_{n}}}-(P\alpha)^{\frac{1}{2}}\langle\Psi_{z},e^{ik_{B_{n}}x}\Psi_{z}\rangle_{E}\right)\left(z_{B_{n}}-\frac{(P\alpha)^{\frac{1}{2}}}{(1-\delta)\langle\Psi_{z},\Psi_{z}\rangle}\langle\Psi_{z},e^{-ik_{B_{n}}x}\Psi_{z}\rangle_{E}\right)\geq0,
\]
completing the square yields
\[
(1-\delta)\langle\Psi_{z},\Psi_{z}\rangle_{E}|z_{B_{n}}|^{2}-\overline{z_{B_{n}}}(P\alpha)^{\frac{1}{2}}\langle\Psi_{z},e^{-ik_{B_{n}}x}\Psi_{z}\rangle_{E}-z_{B_{n}}(P\alpha)^{\frac{1}{2}}\langle\Psi_{z},e^{ik_{B_{n}}x}\Psi_{z}\rangle_{E}
\]
\[
\geq\frac{-(P\alpha)\left|\langle\Psi_{z},e^{-ik_{B_{n}}x}\Psi_{z}\rangle_{E}\right|^{2}}{(1-\delta)\langle\Psi_{z},\Psi_{z}\rangle_{E}}.
\]

The advantage of constructing a Block Hamiltonian in the previous
subsection is that the energy error we incur for disregarding the
``-1'' term in (\ref{eq:10-1}) is proportional to the number of
blocks: $\frac{2K}{P}$, a finite value. 

In the following calculation, in each block $B_{n}$ we choose a mode
$\mathbf{\mathcal{K}_{B_{n}}}$ such that 
\[
\left|\langle\Psi_{z},e^{-i\mathbf{\mathcal{K}_{B_{n}}}x}\Psi_{z}\rangle_{E}\right|^{2}=\underset{k\in B_{n}}{\min}\left|\langle\Psi_{z},e^{-ikx}\Psi_{z}\rangle_{E}\right|^{2}.
\]
As seen in (\ref{eq:a2}) below, we also make use of the continuum
approximation $\sum_{k}\rightarrow L\int dk$ permitted by our model.
We now proceed to extract the Pekar energy functional from (\ref{eq:11-1}):

\[
\sup_{\{k_{B_{n}}\}}\left\langle \Psi,\ H_{K}^{\mbox{Block}}\left(\left\{ k_{B_{n}}\right\} \right)\Psi\right\rangle 
\]

\begin{equation}
\geq\begin{array}[t]{c}
\int\left\langle \Psi_{z},\ \Psi_{z}\right\rangle _{\mbox{E}}\times\left((1-\epsilon)\frac{\left\langle \Psi_{z},\ p^{2}\Psi_{z}\right\rangle _{\mbox{E}}}{\left\langle \Psi_{z},\ \Psi_{z}\right\rangle _{\mbox{E}}}-\frac{\alpha P}{(1-\delta)}\underset{B_{n}}{\sum}\frac{\left|\left\langle \Psi_{z},\ e^{-i\mathbf{\mathcal{K}_{B_{n}}}x}\Psi_{z}\right\rangle _{\mbox{E}}\right|^{2}}{\left|\left\langle \Psi_{z},\ \Psi_{z}\right\rangle _{\mbox{E}}\right|^{2}}+\right.\\
\left.-\alpha^{2}\frac{\left\langle \Psi_{z},\ V(\alpha x)\Psi_{z}\right\rangle _{\mbox{E}}}{\left\langle \Psi_{z},\ \Psi_{z}\right\rangle _{\mbox{E }}}\right)\prod_{B_{m}}dz_{B_{m}}d\overline{z_{B_{m}}}-(1-\delta)\frac{2K}{P}
\end{array}\label{eq:int}
\end{equation}
\[
=\begin{array}[t]{c}
\int\left\langle \Psi_{z},\ \Psi_{z}\right\rangle _{\mbox{E}}\times\left((1-\epsilon)\frac{\left\langle \Psi_{z},\ p^{2}\Psi_{z}\right\rangle _{\mbox{E}}}{\left\langle \Psi_{z},\ \Psi_{z}\right\rangle _{\mbox{E}}}-\frac{\alpha}{(1-\delta)}\frac{1}{L}\underset{B_{n}}{\sum}\frac{(PL)\left|\left\langle \Psi_{z},\ e^{-i\mathbf{\mathcal{K}_{B_{n}}}x}\Psi_{z}\right\rangle _{\mbox{E}}\right|^{2}}{\left|\left\langle \Psi_{z},\ \Psi_{z}\right\rangle _{\mbox{E}}\right|^{2}}+\right.\\
\left.-\alpha^{2}\frac{\left\langle \Psi_{z},\ V(\alpha x)\Psi_{z}\right\rangle _{\mbox{E}}}{\left\langle \Psi_{z},\ \Psi_{z}\right\rangle _{\mbox{E }}}\right)\prod_{B_{m}}dz_{B_{m}}d\overline{z_{B_{m}}}-(1-\delta)\frac{2K}{P}
\end{array}
\]
\begin{equation}
\geq\begin{array}[t]{c}
\int\left\langle \Psi_{z},\ \Psi_{z}\right\rangle _{\mbox{E}}\times\left((1-\epsilon)\frac{\left\langle \Psi_{z},\ p^{2}\Psi_{z}\right\rangle _{\mbox{E}}}{\left\langle \Psi_{z},\ \Psi_{z}\right\rangle _{\mbox{E}}}-\frac{\alpha}{(1-\delta)}\frac{1}{L}\underset{k}{\sum}\frac{\left|\left\langle \Psi_{z},\ e^{-ikx}\Psi_{z}\right\rangle _{\mbox{E}}\right|}{\left|\left\langle \Psi_{z},\ \Psi_{z}\right\rangle _{\mbox{E}}\right|^{2}}+\right.\\
\left.-\alpha^{2}\frac{\left\langle \Psi_{z},\ V(\alpha x)\Psi_{z}\right\rangle _{\mbox{E}}}{\left\langle \Psi_{z},\ \Psi_{z}\right\rangle _{\mbox{E }}}\right)\prod_{B_{m}}dz_{B_{m}}d\overline{z_{B_{m}}}-(1-\delta)\frac{2K}{P}
\end{array}\label{eq:a1}
\end{equation}
\begin{equation}
=\begin{array}[t]{c}
\int\left\langle \Psi_{z},\ \Psi_{z}\right\rangle _{\mbox{E}}\times\left((1-\epsilon)\frac{\left\langle \Psi_{z},\ p^{2}\Psi_{z}\right\rangle _{\mbox{E}}}{\left\langle \Psi_{z},\ \Psi_{z}\right\rangle _{\mbox{E}}}-\frac{\alpha}{(1-\delta)}\int\left(\frac{\left|\left\langle \Psi_{z},\ e^{-ikx}\Psi_{z}\right\rangle _{\mbox{E}}\right|^{2}}{\left|\left\langle \Psi_{z},\ \Psi_{z}\right\rangle _{\mbox{E}}\right|^{2}}\right)dk+\right.\\
\left.-\alpha^{2}\frac{\left\langle \Psi_{z},\ V(\alpha x)\Psi_{z}\right\rangle _{\mbox{E}}}{\left\langle \Psi_{z},\ \Psi_{z}\right\rangle _{\mbox{E }}}\right)\prod_{B_{m}}dz_{B_{m}}d\overline{z_{B_{m}}}-(1-\delta)\frac{2K}{P}
\end{array}\label{eq:a2}
\end{equation}
\begin{equation}
=\begin{array}[t]{c}
\int\left\langle \Psi_{z},\ \Psi_{z}\right\rangle _{\mbox{E}}\times\left((1-\epsilon)\frac{\left\langle \Psi_{z},\ p^{2}\Psi_{z}\right\rangle _{\mbox{E}}}{\left\langle \Psi_{z},\ \Psi_{z}\right\rangle _{\mbox{E}}}-\frac{\alpha}{(1-\delta)}\int\left(\frac{\left|\Psi_{z}\right|^{4}}{\left|\left\langle \Psi_{z},\ \Psi_{z}\right\rangle _{\mbox{E}}\right|^{2}}\right)dx+\right.\\
\left.-\alpha^{2}\frac{\left\langle \Psi_{z},\ V(\alpha x)\Psi_{z}\right\rangle _{\mbox{E}}}{\left\langle \Psi_{z},\ \Psi_{z}\right\rangle _{\mbox{E }}}\right)\prod_{B_{m}}dz_{B_{m}}d\overline{z_{B_{m}}}-(1-\delta)\frac{2K}{P}
\end{array}\label{eq:a3}
\end{equation}
\[
\geq\begin{array}[t]{c}
\int\left\langle \Psi_{z},\ \Psi_{z}\right\rangle _{\mbox{E}}\times(1-\epsilon)\left(\frac{\left\langle \Psi_{z},\ p^{2}\Psi_{z}\right\rangle _{\mbox{E}}}{\left\langle \Psi_{z},\ \Psi_{z}\right\rangle _{\mbox{E}}}-\frac{\alpha}{(1-\delta)(1-\epsilon)}\int\left(\frac{\left|\Psi_{z}\right|^{4}}{\left|\left\langle \Psi_{z},\ \Psi_{z}\right\rangle _{\mbox{E}}\right|^{2}}\right)dx+\right.\\
\left.-\frac{\alpha^{2}}{(1-\epsilon)^{2}(1-\delta)^{2}}\frac{\left\langle \Psi_{z},\ V(\alpha x)\Psi_{z}\right\rangle _{\mbox{E}}}{\left\langle \Psi_{z},\ \Psi_{z}\right\rangle _{\mbox{E }}}\right)\prod_{B_{m}}dz_{B_{m}}d\overline{z_{B_{m}}}-(1-\delta)\frac{2K}{P}
\end{array}
\]
\begin{equation}
\geq\int\left\langle \Psi_{z},\ \Psi_{z}\right\rangle _{\mbox{E}}\left(\frac{\alpha^{2}e(V)}{(1-\epsilon)(1-\delta)^{2}}\right)\prod_{B_{m}}dz_{B_{m}}d\overline{z_{B_{m}}}-(1-\delta)\frac{2K}{P}\label{eq:a4}
\end{equation}
\[
=\frac{\alpha^{2}e(V)}{(1-\epsilon)(1-\delta)^{2}}-(1-\delta)\frac{2K}{P},
\]
where $e(V)$ is the Pekar energy, defined in (\ref{eq:12}). Above
(\ref{eq:a3}) follows from the Plancherel's theorem, and (\ref{eq:a4})
follows from the scaling properties of our Pekar functional.

\subsection{Controlling the Error Terms}

We have the following upper and lower bounds on the ground-state energy:
\begin{equation}
\alpha^{2}e(V)\geq E_{\alpha}(V)\geq\frac{e(V)\alpha^{2}}{(1-\epsilon)(1-\delta)^{2}}-(1-\delta)\frac{2K}{P}-\frac{1}{2}-(\triangle E)-\frac{2\alpha KP^{2}\pi^{2}}{\delta(\triangle E)}\label{dish}
\end{equation}
\[
=\alpha^{2}e(V)-\underbrace{\underset{\mbox{error-term}}{\left(\left(\frac{\alpha^{2}\delta^{2}-2\alpha^{2}\delta-\frac{8\alpha^{3}}{K}+\frac{16\alpha^{3}\delta}{K}-\frac{8\alpha^{3}\delta^{2}}{K}}{1-2\delta+\delta^{2}-\frac{8\alpha}{K}+\frac{16\alpha\delta}{K}-\frac{8\alpha\delta^{2}}{K}}\right)e(V)-(1-\delta)\frac{2K}{P}-\frac{1}{2}-(\triangle E)-\frac{2\alpha KP^{2}\pi^{2}}{\delta(\triangle E)}\right)}},
\]
since we noted, when using an ultraviolet cutoff above, that the parameters
$\epsilon$ and $K$ must satisfy the coupling relation: $\epsilon=\frac{8\alpha}{K}$.
We now choose specific values (in orders of $\alpha$) for the parameters
$K,\delta,P$ and $\triangle E$ so that the error-term above is of
an order less than $\alpha^{2}$, while satisfying the following constraints:
$0<\delta<1$ and $P<K$ when $\alpha\gg1$. In an attempt to make
the error-term as small as possible, we have chosen 
\[
\delta=c_{1}\alpha^{-\frac{1}{7}},K=c_{2}\alpha^{\frac{76}{49}},P=c_{3}\alpha^{\frac{5}{49}}\mbox{\ and\ }\triangle E=c_{4}\alpha^{\frac{64}{49}}.
\]
From (\ref{dish}) we conclude 
\[
\alpha^{2}e(V)\geq E_{\alpha}(V)\geq\alpha^{2}e(V)-C\alpha^{\frac{71}{49}}.
\]

This proves Theorem 1.

\section{Differentiating the Pekar Energy: Proof of Lemma 5}
\begin{proof}
Given any $\epsilon>0$ and for any finite parameter $\delta>0$,
choose $u_{\delta}\in H^{1}(\mathbb{R})$ so that $\|u_{\delta}\|_{2}=1$,
and 
\begin{equation}
\int_{\mathbb{R}}\left(|u_{\delta}'\mid^{2}-V(x)|u_{\delta}|^{2}-|u_{\delta}|^{4}+\delta W(x)|u_{\delta}|^{2}\right)dx-e(V+\delta W)<\epsilon\label{eq:2-1}
\end{equation}
Since $u_{V}$ is not necessarily the minimizer corresponding to the
perturbed Pekar energy $e(V+\delta W)$, 
\begin{equation}
\int_{\mathbb{R}}\left(|u_{\delta}'|^{2}-V(x)|u_{\delta}|^{2}-|u_{\delta}|^{4}+\delta W(x)|u_{\delta}|^{2}\right)dx\leq\int_{\mathbb{R}}\left(|u_{V}'|^{2}-V(x)|u_{V}|^{2}-|u_{V}|^{4}+\delta W(x)|u_{V}|^{2}\right)dx,\label{eq:12-1-1}
\end{equation}
and 
\begin{equation}
\delta\int_{\mathbb{R}}|u_{V}|^{2}W(x)dx+e(V)\geq e(V+\delta W).\label{eq:3-1}
\end{equation}
Since $u_{\delta}$ is not necessarily the minimizer corresponding
to the Pekar energy $e(V)$, 

\begin{equation}
\int_{\mathbb{R}}\left(|\nabla u_{\delta}|^{2}+V(x)|u_{\delta}|^{2}-|u_{\delta}|^{4}+\delta W(x)|u_{\delta}|^{2}\right)dx\geq e(V)+\delta\int_{\mathbb{R}}|u_{\delta}|^{2}W(x)\, dx,\label{eq:4-1}
\end{equation}
and 
\begin{equation}
\int_{\mathbb{R}}|u_{V}|^{2}W(x)dx\geq\frac{e(V+\delta W)-e(V)}{\delta}\geq\int_{\mathbb{R}}|u_{\delta}|^{2}W(x)dx\label{eq:5-111}
\end{equation}
In parallel had we chosen $\delta<0$ we would have arrived at the
relation in (\ref{eq:5-111}) with the inequalities only reversed. 

It suffices to show 
\begin{equation}
\lim_{\delta\rightarrow0}\int_{\mathbb{R}}|u_{\delta}|^{2}W(x)\, dx=\int_{\mathbb{R}}|u_{V}|^{2}W(x)\, dx,\label{eq:6-1}
\end{equation}
where $W(x)$ is a bounded measure on the real line. We show this
by arguing that $u_{\delta}$ converges uniformly to $u_{V}$ on the
real line.

In the limit $\delta\rightarrow0$, the set of functions $\{u_{\delta}\}$
from (\ref{eq:2-1}) above is in fact a minimizing sequence of the
Pekar functional $\mathcal{E}_{V}$ (see (\ref{eq:0})) with the Pekar
energy $e(V)$. For $\delta$ small enough, since $e(V)<0$, 
\[
1\geq1+e(V)\geq\mathcal{E}_{V}(u_{\delta})=\int_{\mathbb{R}}\left(|u_{\delta}^{'}|^{2}-|u_{\delta}|^{4}-V(x)|u_{\delta}|^{2}\right)dx
\]
\begin{equation}
\geq\|u_{\delta}^{'}\|_{2}^{2}-\|u_{\delta}\|_{\infty}^{2}-\|V\|_{\infty}\label{eq:99}
\end{equation}
\begin{equation}
\geq\frac{1}{2}\|u_{\delta}^{'}\|_{2}^{2}-\frac{1}{2}-\|V\|_{\infty}.\label{eq:99.1}
\end{equation}
In (\ref{eq:99}) we use that $V\in C^{1}(\mathbb{R})$ and is a symmetric
decreasing function. In (\ref{eq:99.1}) we use the one-dimensional
Sobolev inequality, whence we have the uniform bound 
\[
\|u_{\delta}^{'}\|_{2}^{2}<3+2\|V\|_{\infty}.
\]

The sequence $\{u_{\delta}\}$ is thus uniformly bounded on $H^{1}(\mathbb{R})$,
and $\{u_{\delta}\}$- or a subsequence thereof- has a weak limit
$u\in H^{1}(\mathbb{R})$. 

Since $V$ is a symmetric decreasing function, using rearrangement
inequalites we see that given any $\epsilon>0$, there is a compact
set $K_{\epsilon}$ and a parameter $D_{\epsilon}>0$ such that 
\begin{equation}
\{x\in\mathbb{R}:\ |u_{\delta}|^{2}\geq\epsilon\ \mbox{for all }0<\delta<D_{\epsilon}\}\subseteq K_{\epsilon}.\label{eq:99.2}
\end{equation}
Since $V$ vanishes at infinity, we see that given any $\epsilon>0$
there is a compact set $K_{\epsilon}$ such that 
\[
\{x\in\mathbb{R}:\ V(x)\geq\epsilon\}\subseteq K_{\epsilon}.
\]
The compact set $K_{\epsilon}$ obviously has finite measure, so appealing
to Theorem 8.6 in {[}LiLo{]} we conclude that the sequence $\{u_{\delta}\}$
converges to $u$ \textit{strongly} in $L^{p}(K_{\epsilon})$ for
all $p\leq\infty$, and that $\{u_{\delta}\}$ converges \textit{pointwise
uniformly} to $u$ on $K_{\epsilon}$. The latter convergence result
follows from the fundamental theorem of calculus. 

Thus,
\[
-\int_{\mathbb{R}}|u_{\delta}|^{2}V(x)dx-\int_{\mathbb{R}}|u_{\delta}|^{4}dx\longrightarrow-\int_{\mathbb{R}}|u|^{2}V(x)dx-\int_{\mathbb{R}}|u|^{4}dx.
\]
This establishes the weak lower semicontinuity of the Pekar functional
$\mathcal{E}_{V}$, and the weak limit $u$ is in fact the minimizer
of $\mathcal{E}_{V}$. But we see from Proposition 3 above that $\mathcal{E}_{V}$
admits a \textit{unique} minimizer $u_{V}$, so $u\equiv u_{V}$.
Thus $u_{\delta}$ converges pointwise uniformly to $u_{V}$ on $K_{\epsilon}$,
and from (\ref{eq:99.2}) we easily see that the convergence can be
made uniform on the entire real line. 

The relation in (\ref{eq:6-1}) now follows for any bounded measure
$W(x)$ on the real line, and the perturbed Pekar energy $e(V+\delta W)$
is indeed differentiable at $\delta=0$: 
\[
\lim_{\delta\rightarrow0}\frac{e(V+\delta W)-e(V)}{\delta}=\int_{\mathbb{R}}|u_{V}|^{2}W(x)dx.
\]

\end{proof}

\section{Proof of Theorem 4}

\noindent In the proof, $\langle\cdot,\cdot\rangle$ should be read
as the inner product on $\mathcal{F}\otimes L^{2}(\mathbb{R})$. 
\begin{proof}
\noindent Recall the perturbed Hamiltonian $H_{\alpha}(V+\delta W)$
defined in (\ref{eq:s1}): 
\[
H_{\alpha}(V+\delta W)=H_{\alpha}(V)+\alpha^{2}\delta W(\alpha x)
\]
and its ground state energy 
\[
E_{\alpha}(V+\delta W)=\alpha^{2}e(V+\delta W)
\]
discussed in (\ref{eq:s2})- (\ref{eq:s4}). 

\noindent Let $\Psi_{\alpha}$ be an approximate ground-state wave
function (see Definition 2 above) of the Hamiltonian $H_{\alpha}(V)$
given in (\ref{eq:9999}). From the variational principle, 

\[
E_{\alpha}(V+\delta W)-\langle\Psi_{\alpha},\ H_{\alpha}(V)\Psi_{\alpha}\rangle\leq\alpha^{2}\delta\langle\Psi_{\alpha},\ W(\alpha x)\Psi_{\alpha}\rangle,
\]
so that 
\[
\alpha^{-2}\left\{ E_{\alpha}(V+\delta W)-\langle\Psi_{\alpha},\ H_{\alpha}(V)\Psi_{\alpha}\rangle\right\} \leq\delta\langle\Psi_{\alpha},\ W(\alpha x)\Psi_{\alpha}\rangle,
\]
and

\[
\liminf_{\alpha\rightarrow\infty}\ \alpha^{-2}\left\{ E_{\alpha}(V+\delta W)-E_{\alpha}(V)\right\} \leq\delta\left(\liminf_{\alpha\rightarrow\infty}\langle\Psi_{\alpha},\ W(\alpha x)\Psi_{\alpha}\rangle\right).
\]
From Theorem 1 and (\ref{eq:s2}) above, 
\[
e(V+\delta W)-e(V)\leq\delta\left(\liminf_{\alpha\rightarrow\infty}\left(\int_{\mathbb{R}}W(x)\left[\frac{1}{\alpha}\|\Psi_{\alpha}\left(\frac{\circ}{\alpha}\right)\|_{\mathcal{F}}^{2}\right]dx\right)\right).
\]
For $\delta>0$ 
\[
\frac{e(V+\delta W)-e(V)}{\delta}\leq\liminf_{\alpha\rightarrow\infty}\int_{\mathbb{R}}W(x)\left(\frac{1}{\alpha}\|\Psi_{\alpha}\left(\frac{\circ}{\alpha}\right)\|_{\mathcal{F}}^{2}\right)dx,
\]
and 
\[
\lim_{\delta\rightarrow0^{+}}\frac{e(V+\delta W)-e(V)}{\delta}\leq\liminf_{\alpha\rightarrow\infty}\left(\int_{\mathbb{R}}W(x)\left[\frac{1}{\alpha}\|\Psi_{\alpha}\left(\frac{\circ}{\alpha}\right)\|_{\mathcal{F}}^{2}\right]dx\right).
\]
From Lemma 5, 
\[
\int_{\mathbb{R}}W(x)\ |u_{V}|^{2}dx\leq\liminf_{\alpha\rightarrow\infty}\left(\int_{\mathbb{R}}W(x)\left[\frac{1}{\alpha}\|\Psi_{\alpha}\left(\frac{\circ}{\alpha}\right)\|_{\mathcal{F}}^{2}\right]dx\right).
\]
For $\delta<0$, we similarly arrive at the following relation: 
\[
\int_{\mathbb{R}}W(x)|u_{V}|^{2}dx\geq\limsup_{\alpha\rightarrow\infty}\left(\int_{\mathbb{R}}W(x)\left[\frac{1}{\alpha}\|\Psi_{\alpha}\left(\frac{\circ}{\alpha}\right)\|_{\mathcal{F}}^{2}\right]dx\right).
\]
We thus conclude: 
\[
\int_{\mathbb{R}}W(x)|u_{V}|^{2}dx=\lim_{\alpha\rightarrow\infty}\left(\int_{\mathbb{R}}W(x)\left[\frac{1}{\alpha}\|\Psi_{\alpha}(\frac{\circ}{\alpha})\|_{\mathcal{F}}^{2}\right]dx\right),
\]
which is our desired convergence relation for any approximate ground-state
wave function with $W$ a bounded measure on the real line. \end{proof}

\end{document}